\documentstyle[12pt,aasms4]{article}

\newcommand{\be}{\begin{equation}}
\newcommand{\ee}{\end{equation}}

\lefthead{Wei \& Lu}
\righthead{Temporal Properties of GRB Afterglow}

\begin{document}

\title{Diverse Temporal Properties of GRB Afterglow}

\author{D.M. Wei}
\affil{Purple Mountain Observatory, Academia Sinica, Nanjing 210008, 
P.R.China} 
\and
\author{T. Lu$^{2,1,3}$}
\affil{$^{1}$Department of Astronomy, Nanjing University, Nanjing 210093, 
P.R. China\\
$^{2}$CCAST(World Laboratory) P.O.Box.8730, Beijing 100080, P.R.China\\
$^{3}$LCRHEA, IHEP, CAS, Beijing, P.R.China\\}
\authoraddr{pmoyl@pub.nj-online.nj.js.cn}

\begin{abstract}

The detection of delayed X-ray, optical and radio emission, "afterglow",
associated with $\gamma$-ray bursts (GRBs) is consistent with fireball
models, where the emission are produced by relativistic expanding blast
wave, driven by expanding fireball at cosmogical distances. 
The emission mechanisms
of GRB afterglow have been discussed by many authors and synchrotron 
radiation is believed to be the main mechanism. The observations show that
the optical light curves of two observed gamma-ray bursts, GRB970228 and
GRB GRB970508, can be described by a simple power law, which seems to support 
the synchrotron radiation explanation. However, here we shall show that under
some circumstances, the inverse Compton scattering (ICS) may play an
important role in emission spectrum and this may influence the temporal
properties of GRB afterglow. We expect that the light curves of GRB 
afterglow may consist of multi-components, which depends on the fireball 
parameters.
\end{abstract}

\keywords{gamma-rays: bursts --- radiation mechanisms: non-thermal}
\newpage

\section{Introduction}

The origin of $\gamma$-ray bursts has been one of the great unsolved 
mysteries in high energy astrophysics for about 30 years. The recent 
discovery of fading sources at X-ray and optical wavelengths coincident
with the locations of some $\gamma$-ray bursts thus provides a good 
opportunity to probe the nature of these high energy events 
(\cite{Costa97,van97,Bond97}). The detection of
absorption lines in the optical afterglow of GRB970508 provides the 
first direct estimate of source distance, constraining the redshift
of GRB970508 to the range of $0.8 \le z \le 2.3$ (\cite{Met97}).

The observed properties of GRB afterglow are broadly consistent with models 
based on relativistic blast waves at cosmological distances
(\cite{MR97,V97,W97a,W97b,WRM97,WL97}).
In fireball models of GRB afterglow, the huge energy released by an
explosion ($\sim 10^{52}\,ergs$) is converted into kinetic energy of a shell 
expanding at ultra-relativistic speed. After the main GRB event occurred,
the fireball continues to propagate into the surrounding gas, driving 
an ultra-relativistic shock into the ambient medium. The expanding 
shock continuesly heats fresh gas and accelerates relativistic electrons
to very  high energy, which produce the delayed emission on timescale
of days to months.

There are only two $\gamma$-ray bursts, GRB970508 and GRB970228, 
related with which afterglows in optical band have been observed.
In the fireball model, it is widely 
believed that afterglow is produced through synchrotron radiation of
relativistic electrons, and the effect of inverse Compton scattering 
can be neglected, which semms to be supported by the observations that  
the light curves of above two gamma-ray bursts can be described by a
simple power law (\cite{Gal97}). However, here we show that under 
some circumstances, the ICS may have important effect on the emission 
spectrum and this will affect the light curve of GRB afterglow.

In next section, we use two different methods to calculate the specific 
intensity of ICS. One is to calculate the emission spectrum directly,
the other is to estimate the relative importance of ICS and synchrotron
radiation by comparing the energy density of magnetic field and synchrotron
radiation. Of course, these two methods give almost the same result. 
In section 3, we discuss the effect of ICS on the temporal behavior of
GRB afterglow, we find that, in general cases, the light curve may 
be diverse. Finally some discussions and conclusions
are given in section 4.

\section{The intensity of inverse Compton scattering}

The process of inverse Compton scattering has also been discussed by several
authors (e.g. \cite{W97a}), but they have not considered the influence
of ICS on GRB afterglow, and they think ICS to be not important in GRB 
afterglow. Here we present a detailed calculation of the intensity
of ICS using two different methods.

First, we calculate the emission spectrum of ICS directly. Assuming that 
after shock acceleration, the electrons have a power law energy 
distribution $N(\gamma)=k_{e}\gamma^{-p}\,(\gamma_{min} \le \gamma \le
\gamma_{max})$, then the photon spectrum of ICS can be expressed as
(\cite{BG70})
\be
\frac{dN}{dtdE}=\frac{3}{8}\sigma_{T}ck_{e}f(p)E^{-(p+1)/2}
\int \epsilon^{(p-1)/2}n(\epsilon)d\epsilon
\ee
where $\sigma_{T}$ is the Thomson cross section, $f(p)=2^{p+3}\frac{p^{2}
+4p+11}{(p+3)^{2}(p+1)(p+5)}$, and $c$ is the speed of light. We assume
that the soft photon spectrum also obeys a power law, $n(\epsilon)=n_{0}
\epsilon^{-\alpha}\,(\epsilon_{1} \le \epsilon \le \epsilon_{2})$.
Note that, in our case, the soft photons are produced
by synchrotron radiation from these same electrons, so $\alpha=(p+1)/2$.
Then from eq.(1) we have
\be
\frac{dN}{dtdE}=\frac{3}{8}\sigma_{T}ck_{e}f(p)n_{0}E^{-(p+1)/2}
\ln(\frac{\epsilon_{2}}{\epsilon_{1}})
\ee
In the fireball model, the electron density in the comoving frame is
$n_{e}=\int N(\gamma)d\gamma=\Gamma n_{1}$, where $n_{1}$ is the density
of surrounding gas and $\Gamma$ is the bulk Lorentz factor, 
then we have $k_{e}=(p-1)\Gamma n_{1}\gamma_{min}^{p-1}$.
Thus we can obtain the ratio of the specific intensity of ICS at peak
energy ($E=\epsilon_{n}=\gamma_{min}^{2}\epsilon_{m}$) to the 
synchrotron radiation at peak energy ($E=\epsilon_{m}$)
\be
y=\frac{I_{ICS}}{I_{syn}}
=\frac{3}{8}\sigma_{T}(p-1)n_{1}f(p)r\ln (\frac{\epsilon_{2}}{\epsilon_{1}})
\ee
where $\ln (\frac{\epsilon_{2}}{\epsilon_{1}})=2\ln (\frac{\gamma_{max}}
{\gamma_{min}})$. The minimum and maximum electron Lorentz factors can
be estimated by physical processes. For a power law energy distribution,
the average electron Lorentz factor $\bar{\gamma} \sim \gamma_{min}$ for
$p >2$, and it is widely believed that after the shock acceleration,
the average electron energy should be $\bar{\gamma} \sim \xi_{e}(m_{p}/
m_{e})\Gamma$, where $m_{p}\,(m_{e})$ is the mass of proton (electron).
The maximum electron energy 
may be limited by shock acceleration and radiative energy loss. It
has been shown that $\gamma_{max} \sim 5\times 10^{7}B^{-1/2}$ 
(\cite{CW96}), where $B$ is the magnetic field strength, which is usually 
written as $B=(\xi_{B}8\pi \Gamma^{2}n_{1}m_{p}c^{2})^{1/2}$, so
$\ln (\frac{\gamma_{max}}{\gamma_{min}}) \sim 10$. In addition, 
after the shock acceleration the spectral index of electron distribution 
$p$ is typically between 2 and 3 (e.g. \cite{BE87}), therefore
we see that $y \simeq 10^{-24}n_{1}r$.

There is another way to estimate the intensity of ICS, i.e. to calculate 
the ratio of the synchrotron radiation energy density ($u_{syn}$) to 
the magnetic energy density ($u_{B}$), $R=u_{syn}/u_{B}$, $u_{B}=
B^{2}/8\pi$ and $u_{syn} \sim n_{e}P_{syn}r/\Gamma c$, where $P_{syn}$
is the synchrotron radiation power. This method is simpler and more
important, since the value of $R$ indicates which process is more
efficient for electron energy loss, inverse Compton scattering or 
synchrotron radiation, so it is necessary to calculate the value of $R$
carefully. It is easy to show that $R \simeq 10^{-24}\bar{\gamma}^{2}
n_{1}r$. In the afterglow model, if the fireball expands outward
adiabatically, we have 
\be
r/r_{0}=(t/t_{0})^{1/4} \hspace{25mm}\Gamma/\Gamma_{0}=(t/t_{0})^{-3/8}
\ee
where the deceleration radius $r_{0}=r_{d}=5\times 10^{16}(E_{51}/n_{1})^
{1/3}(\Gamma_{0}/100)^{-2/3}\,(cm)$, and the characteristic 
time $t_{0}\simeq r_{0}/2\Gamma_{0}^{2}c
\simeq 10^{2}(E_{51}/n_{1})^{1/3}(\Gamma_{0}/100)^{-8/3}\,(s)$ 
(\cite{MR97}) with $E_{51}$ being the burst
energy in units of $10^{51}\,ergs$ and $\Gamma_{0}$ the initial 
Lorentz factor. Then we obtain
\be
R=17\xi_{e}^{2}n_{1}^{1/2}E_{51}^{1/2}t_{day}^{-1/2}
\ee
Obviously, the emission power of ICS is not much smaller than
that of synchrotron radiation even for several days after the GRB
event, thus the effect of ICS should not be neglected. The ratio of 
intensities of ICS to synchrotron radiation at peak energy is 
$I_{ICS}/I_{syn}\sim R\epsilon_{syn}/\epsilon_{ICS}\sim 
R\bar{\gamma}^{-2}\sim 10^{-24}n_{1}r$, which is 
in agreement with the previous result.

\section{The effect of ICS on GRB afterglow}

In the case where the effect of ICS cannot be neglected, the emission 
spectrum should consist of two components. There exists a critical 
energy $\epsilon_{c}$, below which the spectrum is dominated by synchrotron
radiation and above which the spectrum is dominated by ICS, our
purpose is to calculate the value of $\epsilon_{c}$.

We assume that, in the comoving frame, the synchrotron radiation intensity
has the form $I_{\epsilon} \propto \epsilon^{-\alpha}$ for $\epsilon 
<\epsilon_{m}$ and $I_{\epsilon} \propto \epsilon^{-\beta}$ for 
$\epsilon >\epsilon_{m}$. Since in our situation
the soft photons produced through synchrotron radiation are scattered by
the same electrons, so the Compton scattered spectrum should have nearly 
the same  form as that of synchrotron radiation, i.e. 
$I_{\epsilon} \propto \epsilon^{-\alpha}$ for $\epsilon <\epsilon_{n}$ and
$I_{\epsilon} \propto \epsilon^{-\beta}$ for $\epsilon >\epsilon_{n}$, 
therefore the total intensity is $I_{\epsilon} \propto \epsilon^{-\alpha}$
for $\epsilon <\epsilon_{m}$ or $\epsilon _{c} <\epsilon< \epsilon_{n}$,
and $I_{\epsilon} \propto \epsilon^{-\beta}$ for $\epsilon_{m} < \epsilon
< \epsilon_{c}$ or $\epsilon > \epsilon_{n}$. 
Then from eq.(4) and the relation
$I_{\epsilon_{m}}(\frac{\epsilon_{c}}{\epsilon_{m}})^{-\beta}=I_{\epsilon_
{n}}(\frac{\epsilon_{c}}{\epsilon_{n}})^{-\alpha}$ we can obtain the
value of $\epsilon_{c}$
\be
\epsilon_{c}=f(\alpha,\beta)\xi_{B}^{1/2}\xi_{e}^{2-\frac{2\alpha}{\beta-
\alpha}}n_{1}^{-\frac{3-\alpha}{4(\beta-\alpha)}}E_{51}^{\frac{1}{2}-
\frac{1+\alpha}{4(\beta-\alpha)}}t_{day}^{-\frac{3}{2}-\frac{1-3\alpha}
{4(\beta-\alpha)}} \hspace{15mm} eV
\ee
where $f(\alpha,\beta)=20\times 17^{-\frac{1}{\beta-\alpha}}10^{\frac
{6(1-\alpha)}{\beta-\alpha}}8^{\frac{2(1-\alpha)}{\beta-\alpha}}$.

It has been shown that, in the adiabatic case, the electron Lorentz
factor $\gamma_{e}\propto t^{-3/8}$, the typical energy $\epsilon_{m}
\propto t^{-3/2}$, and the comoving specific intensity of synchrotron
radiation at peak energy is $I_{\epsilon_{m}}'\propto t^{-1/8}$ 
(\cite{MR97}). From the relation $I_{ICS}/I_{syn} \sim R\gamma_{e}^{-2}$
it is easy to show that the intensity of ICS at peak energy 
$I_{\epsilon_{n}}'\propto t^{1/8}$ and $\epsilon_{n} \propto t^{-9/4}$,
then the observed peak flux $F_{\epsilon_{m}}\propto t^{2}\gamma^{5}
I_{\epsilon_{m}}'\propto t^{0} \sim$ constant, and $F_{\epsilon_{n}}
\propto t^{2}\gamma^{5}I_{\epsilon_{n}}' \propto t^{1/4}$. Therefore
we can conclude that, if our observation is fixed at energy $\epsilon$,
then the observed flux $F_{\epsilon}\propto F_{\epsilon_{m}}(\frac
{\epsilon}{\epsilon_{m}})^{-\alpha}\propto t^{-\frac{3}{2}\alpha}$ for
$\epsilon <\epsilon_{m}$,
$F_{\epsilon}\propto F_{\epsilon_{m}}(\frac
{\epsilon}{\epsilon_{m}})^{-\beta}\propto t^{-\frac{3}{2}\beta}$ for
$\epsilon_{m} <\epsilon <\epsilon_{c}$, 
$F_{\epsilon}\propto F_{\epsilon_{n}}(\frac
{\epsilon}{\epsilon_{n}})^{-\alpha}\propto t^{\frac{1}{4}-\frac{9}{4}\alpha}$ 
for $\epsilon_{c} <\epsilon <\epsilon_{n}$, and
$F_{\epsilon}\propto F_{\epsilon_{n}}(\frac
{\epsilon}{\epsilon_{n}})^{-\beta}\propto t^{\frac{1}{4}-\frac{9}{4}\beta}$ 
for $\epsilon >\epsilon_{n}$. 

Here the most interesting quantity is the critical energy $\epsilon_{c}$, 
which is dependent on the fireball parameters, i.e. the fireball energy,
surrounding gas density, energy fractions in electrons and magnetic field,
and the spectral index of synchrotron radiation. In particular, it is easy 
to show that the effect of inverse Compton scattering is important only for 
large values of spectral index. As an example, we take
$\alpha=0.25,\,\beta=1.4$ (these values are consistent with the observed 
$\gamma$-ray burst spectra), then the value of $\epsilon_{c}$
\be
\epsilon_{c}=1.85(\frac{\epsilon_{syn}}{100KeV})(\frac{\gamma_{0}}{600})^
{-4}\xi_{e}^{-0.43}n_{1}^{-1.1}E_{51}^{0.23}(\frac{t}{6{\rm day}})^{-1.55}
\hspace{12mm} eV
\ee
where we have used the relation $\epsilon_{syn}=5.2\times 10^{-3}
\gamma_{0}^{4}\xi_{B}^{1/2}\xi_{e}^{2}n_{1}^{1/2}\,eV$, which is
the peak energy of $\gamma$-ray burst spectrum. So we can see that,
for the typical values of GRB events, i.e. $E \sim 10^{51} \,ergs,\,
n \sim 1\,cm^{-3},\,\xi_{e} \sim 0.3$, the critical energy $\epsilon_{c}$
crosses the optical band for about six days after the burst.
Furthermore, we can calculate the peak energy of ICS
\be
\epsilon_{n}=1.87(\frac{\epsilon_{syn}}{100KeV})(\frac{\gamma_{0}}{600})^
{-4}(\frac{\xi_{e}^{2}}{0.1})n_{1}^{-3/4}E_{51}^{3/4}(\frac{t}{60{\rm day}})
^{9/4} \hspace{12mm} eV 
\ee
Obviously, if we take the same parameters as above, then 
$\epsilon_{n}$ should cross the optical band for about two months,
so we expect that the time during which the optical flux varies as
$F \propto t^{\frac{1}{4}-\frac{4}{9}\alpha}$ rather than $F \propto 
t^{-\frac{3}{2}\beta}$, and according to the values of $\alpha$ and $\beta$,
the brightness may decrease very slowly than before.

\section{Discussion and conclusion}

The detection of $\gamma$-ray burst in the optical and radio bands 
has greatly furthered our understanding these objects, especially
the shape of the light curves of GRB afterglow provide important 
information on exploring their emission mechanisms. Here we present 
a detailed calculation to show that the temporal properties of GRB
afterglow may be quite diverse due to the effect of inverse Compton 
scattering.

We have shown that the inverse Compton scattering may play an important
role in the GRB afterglow. From eq.(5) one sees that $R\propto t^{-1/2}$,
it seems that ICS is more efficient than synchrotron radiation in the
early time of GRB afterglow (especially in $\gamma$-ray burst epoch,
it seems that ICS dominates the electron energy loss), however, in fact this
is not true, since in the early time, the photon energy of synchrotron
radiation is so high that in the electron rest frame the photon energy 
is greatly larger than $m_{e}c^{2}$, so the Compton scattering occurs
in the extreme Klein-Nishina limit, and thus the scattering is not 
efficient. It can easily be shown that Compton scattering is efficient 
only when about $t>10^{3}\,s$ after GRB event.

Here we point out that the light curves of GRB afterglow may 
consist of four components, which depend on the characteristic energies
$\epsilon_{m},\,\epsilon_{c},\,\epsilon_{n}$ and the detector frequency
$\epsilon$. However, it should be noted that not all the GRB afterglows
could contain so many components, the necessary condition for having 
four components is $\epsilon_{c}< \epsilon_{n}$. It is obvious that 
if $\epsilon_{c}>\epsilon_{n}$, the light curves should have three
components. In addition, if the ICS efficiency is low enough
that there is no intersection between the spectrum of ICS and synchrotron
radiation, then the light curves only have two components. 

It should be noted that the effect of inverse Compton scattering is
strongly dependent on the model parameters, especially on the spectral 
index of synchrotron radiation, the contribution from inverse Compton 
scattering is important only for those bursts with large spectral indices,
and the spectral indices may be estimated by the ratio of X-ray flux
to the optical flux.

The observations from BATSE show that the spectra of GRBs are rather
diverse, varying from burst to burst. For most bursts, their spectra
can be described by a broken power law, and the distribution of the 
spectral indices below and above the peak energy is rather wide. In
general, below peak energy the spectral indices ($\alpha$) are mainly 
between  0 and 0.5, while above peak energy the slopes ($\beta$) are 
typically between 1 and 1.5 (\cite{Band93}), so we expect that the 
temporal properties of GRB afterglows (which strongly depend on the 
parameters $\alpha$ and $\beta$) would be very diverse, and
contain multi-components.

\acknowledgements{We thank the referee for several important comments
which improved this paper. This work was supported by the
National Natural Science Foundation and the National Climbing 
Project on Fundamental Researches of China.}

\end{document}